\documentclass[aps,prb,onecolumn,superscriptaddress,amsmath,amssymb]{revtex4-2}
\pdfoutput=1
\usepackage[english]{babel}
\usepackage{amsthm}
\usepackage{color}
\usepackage{tikz}
\usepackage{pgfplots}
\pgfplotsset{compat=1.17}
\usepackage{tikz-3dplot}
\usetikzlibrary{shadows}
\usetikzlibrary{intersections,positioning}
\usetikzlibrary{arrows.meta}
\usepackage{epsfig}
\usepackage{graphicx}
\usepackage{bm}
\usepackage{times}
\usepackage{subfigure}
\usepackage{amsfonts}
\usepackage[colorlinks = true,
linkcolor = blue,
urlcolor  = blue,
citecolor = blue,
anchorcolor = blue]{hyperref}
\usepackage{mathtools}
\usepackage{empheq}

\global\long\def\bra#1{\langle #1 |}
\global\long\def\ket#1{| #1 \rangle }
\global\long\def\kket#1{|| #1 \rangle\rangle}

\global\long\def\braket#1#2{\left\langle #1|#2\right\rangle }

\global\long\def\no{\nonumber}

\global\long\def\l{\lambda}

\newcommand{\ir}{\mathrm{i}}
\newcommand{\eE}{\mathrm{e}}
\newcommand{\mM}{{\mathcal{M}}}
\newcommand{\tM}{\widetilde{M}}

\theoremstyle{remark}
\newtheorem*{rem}{Remark}

\begin{document}

\title{Bethe ansatz approach for the steady state of the asymmetric simple exclusion process with open boundaries}
\author{Xin Zhang}
\thanks{Email: xinzhang@iphy.ac.cn}
\affiliation{Beijing National Laboratory for Condensed Matter Physics, Institute of Physics, Chinese Academy of Sciences, Beijing 100190, China}
\author{Fa-Kai Wen}
\affiliation{College of Physics and Electronic Information, Yunnan Normal University, Kunming 650500, China}

\begin{abstract}
We study the asymmetric simple exclusion process with non-diagonal boundary terms under a specific constraint. A symmetric chiral basis is constructed and a special string solution of the Bethe ansatz equations corresponding to the steady state is presented. Using the coordinate Bethe ansatz method, we derive a concise expression for the steady state. The current and density profile in the steady state are also studied.   
\end{abstract}

\maketitle

\section*{Introduction}
The asymmetric simple exclusion process (ASEP) \cite{Spitzer,Derrida1998}, which describes the asymmetric diffusion of hard-core particles with anisotropic hopping rates, is a fundamental model in the study of non-equilibrium statistical mechanics and stochastic processes. The ASEP has been widely used in the research on mRNA translation \cite{Zur2016}, vehicular traffic \cite{Schadschneider2010} and motor-protein transport \cite{Neri2013}.

The ASEP is an integrable system \cite{Essler1996,Schutz2001,OpenASEPjan}. Various techniques such as the matrix product ansatz \cite{Derrida1993,Golinelli2006,Lazarescu2013,Bryc2019} and the Bethe ansatz \cite{Derrida1999,Simon2009,Wen2015,Zhang2019} have be employed to obtain its exact solutions.
The ASEP with generic open boundaries can be mapped to an open XXZ chain with constrained non-diagonal boundary fields through a similarity transformation \cite{Essler1996}. The non-diagonal boundaries break the $U(1)$ symmetry of both the open XXZ spin chain and the open ASEP, which prevents the use of conventional Bethe ansatz methods. However, the open XXZ model with the specific constraint exhibits an alternative symmetry. In Ref. \cite{Chernyak2023}, the authors mapped it to another open XXZ chain which is $U\mathfrak{_q sl}_2$-invariant using representation theory of the two-boundary Temperley-Lieb algebra. It has been shown that the spectrum of this specific constrained open XXZ chain (or open ASEP) can still be parameterized using Baxter's homogeneous $T-Q$ relation \cite{Nepomechie2003, Cao2003,OpenASEPjan,Simon2009}. Consequently, a set of Bethe ansatz equations (BAEs) is derived, which enables the study of the model's physical properties \cite{OpenASEPjan,Jan2006,Jan2011}.

For the open ASEP, the eigenstates of the Markov transition matrix (or the transfer matrix) can be constructed using the modified algebraic Bethe ansatz method introduced in Ref. \cite{Cao2003},  the (modified) coordinate Bethe ansatz proposed in Refs. \cite{Simon2009, Crampe2010}, and the subsequent analogue — the chiral coordinate Bethe ansatz, presented in Refs. \cite{CCBA, PhantomBetheRoots}. However, due to the non-Hermitian nature of the system, constructing all the eigenstates of the Markov transition matrix is challenging. For example, the steady state is represented by a tensor product bra vector $\bra{\Phi}$, which serves as the pseudo-vacuum in the modified Bethe ansatz method and corresponds to a trivial $T-Q$ relation without $Q$-function. In contrast, the ket vector $\ket{\Phi}$ is not factorized and has a notably complex expression.

In this paper, we study the open ASEP with a specific constraint (\ref{constraint}). This constraint exhibits the following features: (1) it enables the construction of additional homogeneous $T-Q$ relations and BAEs; (2) it allows us to introduce a closed set of factorized chiral vectors; (3) it facilitates the parameterization of the eigenstates of the Markov transition matrix, including the right steady state $\ket{\Phi}$, using the solutions of the BAEs and the chiral basis. 
In this paper, we use the coordinate Bethe ansatz approach presented in Refs. \cite{PhantomBetheRoots,CCBA} to construct the steady state. 
We prove analytically that the steady state corresponds to a special string solution of BAEs and it allow us to get an elegant expression for the steady state. The current and density profile of the system under the steady state are also studied analytically and numerically in this paper.  

The paper is organized as follows: Section \ref{sec;Markov} provides a brief introduction to the ASEP with open boundaries. In Section \ref{sec;integrability}, we demonstrate the system's integrability. The $T-Q$ relations are presented in Section \ref{sec;TQ1}. In Section \ref{sec;TQ2}, we introduce the constraint we are interested in and demonstrate another homogeneous $T-Q$ relations. A set of chiral vectors is constructed in Section \ref{sec;Basis}, where we also introduce the chiral coordinate Bethe ansatz method and a special string solution. The expression for the steady state is derived in Section \ref{sec;SteadyState}. Finally, Section \ref{sec;GenericASEP} explores the ASEP with generic open boundaries. Some numerical results are provided in the appendix.

\section{Markov transition matrix}\label{sec;Markov}
\begin{figure}[htbp]
\centering
\begin{tikzpicture}
\draw[color=gray,line width=0.5pt] (0.5,0) --(2.9,0);
\draw[color=gray,line width=0.5pt] (3.1,0) --(3.9,0);
\draw[color=gray,line width=0.5pt] (4.1,0) --(6.5,0);
\draw[fill] (1,0) circle (.1);
\draw[fill] (2,0) circle (.1);
\draw (3,0) circle (.1);
\draw (4,0) circle (.1);
\draw[fill] (6,0) circle (.1);
\draw[fill] (5,0) circle (.1);
\draw[->] (2.0,-0.15) arc (-160 : -20 : 0.5);
\coordinate[label=below:$\frac{q}{q+1}$] (q) at (2.5,-0.65);
\draw[<-] (4.05,-0.15) arc (-160 : -20 : 0.5);
\coordinate[label=below:$\frac{1}{q+1}$] (1) at (4.5,-0.65);
\end{tikzpicture}\\
\begin{tikzpicture}
\draw[color=gray,line width=0.5pt] (0,0) --(0.5,0);
\draw[color=gray,line width=0.5pt] (0.7,0) --(1,0);
\draw[color=gray,line width=0.5pt] (0,-0.4) --(0,0.4);
\draw[fill] (-0.6,0) circle (.1);
\draw (0.6,0) circle (.1);
\draw[->] (-0.57,-0.15) arc (-160 : -20 : 0.6);
\coordinate[label=below:$\alpha$] (a) at (0,-0.7);
\end{tikzpicture}\qquad \quad
\begin{tikzpicture}
\draw[color=gray,line width=0.5pt] (0,0) --(1,0);
\draw[color=gray,line width=0.5pt] (0,-0.4) --(0,0.4);
\draw[fill] (0.6,0) circle (.1);
\draw (-0.6,0) circle (.1);
\draw[<-] (-0.57,-0.15) arc (-160 : -20 : 0.6);
\coordinate[label=below:$\gamma$] (a) at (0,-0.7);
\end{tikzpicture}\\
\begin{tikzpicture}
\draw[color=gray,line width=0.5pt] (0,0) --(-0.5,0);
\draw[color=gray,line width=0.5pt] (-0.7,0) --(-1,0);
\draw[color=gray,line width=0.5pt] (0,-0.4) --(0,0.4);
\draw (-0.6,0) circle (.1);
\draw[fill] (0.6,0) circle (.1);
\draw[<-] (-0.57,-0.15) arc (-160 : -20 : 0.6);
\coordinate[label=below:$\delta$] (a) at (0,-0.7);
\end{tikzpicture}\qquad \quad
\begin{tikzpicture}
\draw[color=gray,line width=0.5pt] (0,0) --(-1,0);
\draw[color=gray,line width=0.5pt] (0,-0.4) --(0,0.4);
\draw (0.6,0) circle (.1);
\draw[fill] (-0.6,0) circle (.1);
\draw[->] (-0.57,-0.15) arc (-160 : -20 : 0.6);
\coordinate[label=below:$\beta$] (a) at (0,-0.7);
\end{tikzpicture}
\caption{The transition rates of ASEP with open boundaries. Here, $\bullet$ and $\circ$ represent occupied sites and empty sites, respectively.}\label{Fig;ASEP}
\end{figure}
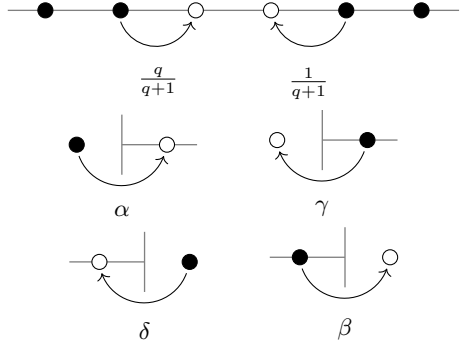

We study the ASEP with open boundaries which is described by the following Markov transition matrix \cite{OpenASEPjan}
\begin{align} 
\mathcal{M}=\mathcal M_1+\sum_{k=1}^{N-1}\mathcal{M}_{k,k+1}+\mathcal{M}_N,\label{MarkovMatrix}
\end{align} 
where
\begin{align}
&\mathcal{M}_{k,k+1}=\left(
\begin{array}{cccc}
	0 & 0 & 0 & 0 \\[4pt]
	0 & -\frac{1}{q+1} & \frac{q}{q+1} & 0 \\[4pt]
	0 & \frac{1}{q+1} & -\frac{q}{q+1} & 0 \\ [4pt]
	0 & 0 & 0 & 0
\end{array}
\right),\qquad \mathcal{M}_1=\left(\begin{array}{cc}
	-\alpha & \gamma \\
	\alpha & -\gamma
\end{array}
\right),\qquad \mathcal{M}_N=\left(\begin{array}{cc}
	-\delta & \beta \\
	\delta & -\beta
\end{array}
\right).
\end{align}
Here $\alpha,\,\beta,\,\gamma,\,\delta$ and $q$ are all non-negative real numbers, and the subscripts indicate on which sites the matrices $\mathcal{M}_{k,k+1}$ and $\mathcal{M}_{j}$ act non-trivially. The transition rates of open ASEP are illustrated in Figure \ref{Fig;ASEP}.

The Markov transition matrix $\mathcal M$ can map to the Hamiltonian of an open spin-$\frac12$ XXZ model with constrained boundary fields through a similarity transformation \cite{Essler1996}
\begin{align} 
\mathcal M=\frac{\sqrt{q}}{2(1+q)}\,G^{-1}HG,\quad G=\bigotimes_{n=1}^N\left(\begin{array}{cc}
1 & 0 \\
0 & \rho q^{\frac{1-n}{2}}
\end{array}\right),\label{Gauge}
\end{align}
where the gauge parameter $\rho$ can be an arbitrary non-zero complex number.
The Hamiltonian $H$ in Eq. (\ref{Gauge}) reads
\begin{align}
H=&\sum_{j=1}^{N-1}\left[\sigma^x_j\sigma_{j+1}^x+\sigma^y_j\sigma_{j+1}^y+\frac{q+1}{2\sqrt{q}}\left(\sigma_j^z\sigma_{j+1}^z-\mathbb{I}\right)\right]+\vec{h}_1\cdot \vec{\sigma}_1+\vec{h}_N\cdot\vec{\sigma}_N\no\\
&+\frac{q-1}{2\sqrt{q}}(\sigma_1^z-\sigma_N^z)-(q+1)\left(\frac{\alpha+\gamma}{\sqrt{q}}\!+\!\frac{\beta+\delta}{\sqrt{q}}\right),\label{H_XXZ}
\end{align}
where $\vec{h}_1$ and $\vec{h}_N$ are defined by 
\begin{align} 
\begin{aligned}
&\vec{h}_1=(q+1)\left(\frac{\gamma+\alpha\rho^2}{\sqrt{q}\rho},\,{\rm i} \frac{\gamma-\alpha\rho^2}{\sqrt{q}\rho},\,\frac{\gamma-\alpha}{\sqrt q}\right),\\
& \vec{h}_N=(q+1)\left(\frac{\beta q^{\frac {N-1}{2}}+\delta\rho^2q^{\frac{1-N}{2}}}{\sqrt{q}\rho},\,{\rm i} \frac{\beta q^{\frac{N-1}{2}}-\delta\rho^2q^{\frac{1-N}{2}}}{\sqrt{q}\rho},\,\frac{\beta-\delta}{\sqrt q}\right).
\end{aligned}
\end{align}
\begin{rem}
For a generic open XXZ chain ($J_{x,y}=1$) with boundary fields, the Hamiltonian should have seven free parameters, e.g., the anisotropic parameter and the components of the two boundary magnetic fields. We observe that the Hamiltonian $H$ given by Eq. (\ref{H_XXZ}) contains only six free parameters: \(q\), \(\alpha, \beta, \gamma, \delta\), and \(\rho\). Therefore, the XXZ model introduced in this paper is not the most generic one, and there exists a constraint.
\end{rem}

As a stochastic process, the left steady state of Markov transition matrix is a simple factorized state \cite{Derrida1993}
\begin{align}
\bra{\Phi}=\bigotimes_{n=1}^N(1,\,1).\label{LeftSS}
\end{align} 

Both $\mathcal{M}$ in (\ref{MarkovMatrix}) and $H$ in (\ref{H_XXZ}) are non-Hermitian operators, and the conjugate transpose of $\bra{\Phi}$ in Eq. (\ref{LeftSS}) is not an eigenstate of $\mathcal{M}$. In this paper, we will use the Bethe ansatz method to study the right steady state of the open ASEP.

\section{Integrability}\label{sec;integrability}

The ASEP is an integrable system and the corresponding $R$-matrix is \cite{Crampe2014}
\begin{align}
&R(x)=\left(
       \begin{array}{cccc}
         a(x) & 0 & 0 & 0 \\
         0 & b^-(x) & c^+(x) & 0 \\
         0 & c^-(x) & b^+(x) & 0 \\
         0 & 0 & 0 & a(u) \\
       \end{array}
     \right),\label{def-R}
\end{align}
where $x$ is the spectral parameter and some functions in the $R$-matrix are defined as follows
\begin{align}
&a(x)=q-x,\quad b^+(x)=q(1-x),\quad b^-(x)=1-x,\no\\
&c^+(x)=(q-1)x,\quad c^-(x)=q-1.\label{Dabc}
\end{align}
The $R$-matrix in (\ref{def-R}) satisfies the Yang-Baxter equation (YBE) \cite{Korepin1997}
\begin{align}
R_{1,2}(x/y)R_{1,3}(x/z)R_{2,3}(y/z)
=R_{2,3}(y/z)R_{1,3}(x/z)R_{1,2}(x/y).\label{YBE}
\end{align}
In order to present the reflection matrices more concisely, we introduce a new set of parameters
$\{a,\,b,\,c,\,d\}$ where
\begin{align}
\begin{aligned}
&\alpha=\frac{(q-1)ac}{(a+1)(c+1)(q+1)},\quad \gamma=\frac{(1-q)}{(a+1)(c+1)(q+1)},\\
&\beta=\frac{(q-1)bd}{(b+1)(d+1)(q+1)},\quad \delta=\frac{(1-q)}{(b+1)(d+1)(q+1)}.
\end{aligned}
\end{align}
The boundary $K$-matrices $K^{\pm}(x)$ are given by \cite{Crampe2014}
\begin{align}
&K^-(x)=\left(
\begin{array}{cc}
(ac+1)x^2+(a+c)x & x^2-1\\
ac(1-x^2) & (ac+1)+(a+c)x \\
\end{array}
\right),\\
&K^+(x)=\left(
\begin{array}{cc}
(bd+1)q^2+(b+d)qx & bd(q^2-x^2) \\
q(x^2-q^2) & (bd+1)qx^2+(b+d)q^2x \\
\end{array}
\right),
\end{align}
which satisfy the following reflection equation (RE)
and dual RE respectively
\begin{align}
R_{1,2}(xy^{-1})K_1^-(x)R_{2,1}(yx) K_2^-(y)=K_2^-(y)R_{1,2}(xy) K_1^-(x)R_{2,1}(y^{-1}x),\label{RE}\\
R_{1,2}(xy^{-1})K_1^+(y)\widetilde R_{1,2}(yx)K_2^+(x)
=K_2^+(x)\widetilde R_{2,1}(yx)K_1^+(y)R_{2,1}(xy^{-1}),\label{DRE}
\end{align}
where
\begin{align}
\widetilde{R}_{1,2}(x)=(R^{t_1}_{1,2}(x))^{-1})^{t_1}.
\end{align}
Then, the transfer matrix can be constructed
\begin{align}
&\tau(x)=\mathrm{tr}_0\left\{K_0^+(x)R_{0,N}(x)\cdots R_{0,1}(x)K_0^-(x)
R_{1,0}(x)\cdots R_{N,0}(x)\right\}.\label{TransferMatrix}
\end{align}

Using the YBE (\ref{YBE}), RE (\ref{RE})
and dual RE (\ref{DRE}), we can prove that the transfer matrix $\tau(x)$ forms a commuting family, i.e., $[\tau(x),\,\tau(y)]=0$.
The Markov transition matrix $\mathcal M$ can be obtained from the transfer matrix as follows
\begin{align}
\mathcal{M}=\left.\frac{1-q}{2(1+q)}\frac{\partial \ln\tau(x)}{\partial x}\right|_{x=1}-\mbox{const}.\label{MM}
\end{align}

\section{\texorpdfstring{$T-Q$ relation}{T-Q relation}}\label{sec;TQ1}
For the ASEP with generic open boundary conditions, the Hilbert space of the system splits into two invariant subspaces, i.e., $\mathcal{H}=\mathcal{H}_1\oplus\mathcal{H}_2$, whose dimensions are ${\rm dim}\,\mathcal{H}_1=2^N-1$ and ${\rm dim}\,\mathcal{H}_2=1$ respectively. The full spectrum of the quantum transfer matrix can be described by two homogeneous $T-Q$ relations, with the number of Bethe roots being $N-1$ and 0 respectively \cite{Cao2003,Nepomechie2003,OpenASEPjan}. Let us recall these two $T-Q$ relations in this section.

First, we introduce some functions
\begin{align}
& a_1(x)=\frac{q^3-qx^2}{q-x^2}
(ax+1)(cx+1)(bx+1)(dx+1)a^{2N}(x),\label{a1}\\
& d_1(x)=\frac{1-x^2}
{q-x^2}(x+aq)
(x+cq)(x+bq)(x+dq)
\left[b^-(x)b^+(x)\right]^N,\label{d1}\\
&a_2(x)=\frac{q^3-qx^2}{q-x^2}
(x+a)(x+c)(x+b)(x+d)
a^{2N}(x),\label{a2}\\
&d_2(x)=\frac{1-x^2}{q-x^2}
(ax+q)(cx+q)(bx+q)(dx+q)
\left[b^-(x)b^+(x)\right]^N.\qquad\label{d2}
\end{align}
\textit{$T-Q$ relation I: ~~~}
The eigenvalue of the transfer matrix $\tau(x)$ can be parameterized by the following $T-Q$ relation
\begin{align}
&\Lambda_1(x)= a_1(x)\,\frac{q^{1-N}Q_1(qx)}{Q_1(x)}
+ d_1(x)\,\frac{q^{N-1}Q_1(q^{-1}x)}{Q_1(x)},\label{TQ-1}
\end{align}
where 
\begin{align}
Q_1(x)=\prod_{k=1}^{N-1}(\l_k-x)(\l_kx-q).\label{Q-1}
\end{align}
The Bethe roots $\{\l_j\}$ are given by the following BAEs
\begin{align}
\frac{d_1(\l_j)}{a_1(\l_j)}=-\prod_{k=1}^{N-1}\frac{(\l_k-q\l_j)(\l_k\l_j-1)}
{(\l_k-q^{-1}\l_j)(\l_k\l_j-q^2)},\quad j=1,\ldots,N-1.\label{BAEs-GC}
\end{align}
The eigenvalue of the Markov transition matrix $\mM$ in terms of $\{\l_j\}$ is
\begin{align}
E=\frac{(1-q)^2}{q+1}\sum_{k=1}^{N-1}\frac{\l_k}
{(\l_k-1)(\l_k-q)}
-(\alpha+\beta+\gamma+\delta).\label{E-lambda1}
\end{align}

The numerical results for small-size systems in Table \ref{Lambda01-com} indicate that the solution of the BAEs in (\ref{BAEs-GC}) yields $2^N-1$ eigenvalues of $\mM$, leaving the energy level $E=0$ unaccounted for.

\textit{$T-Q$ relation II: ~~~} There exists second homogeneous $T-Q$ relation
\begin{align}
&\Lambda_2 (x)=a_{2}(x)+d_{2}(x),\quad\label{TQ-2}
\end{align}
which only gives the trivial eigenvalue of the Markov transition matrix $E=0$.

The functions $\Lambda_1(u)$ and 
$\Lambda_2(u)$ correspond to $\mathcal{H}_1$ and $\mathcal{H}_2$, respectively. The right eigenstates of $\tau(x)$ belonging to $\mathcal{H}_1$ and the left eigenstates belonging to $\mathcal{H}_2$
can be constructed via either the modified algebraic Bethe ansatz method \cite{Cao2003} or the (chiral) coordinate Bethe ansatz method \cite{Simon2009,Crampe2014,CCBA}.

In the generic case, constructing the dual eigenstates, namely the left eigenstates belonging to $\mathcal{H}_1$ and the right eigenstates belonging to $\mathcal{H}_2$, is problematic.
However, under certain constraint, the Hilbert space exhibits another subspaces, which allow us to use the Bethe ansatz method to construct the ``missing" eigenstates of $\mM$, such as the right steady state $\ket{\Phi}$. This inspiring discovery motivates our research.

\section{\texorpdfstring{Another $T-Q$ relations under certain constraint}{Another T-Q relations under certain constraint}}\label{sec;TQ2}

In this paper, we study the open ASEP with the following constraint \cite{Crampe2010}
\begin{align}
abcdq^{N-1-M}=1, \quad \mbox{or equivalently}\,\, \alpha\beta q^{N-1-M}=\gamma\delta,\quad M=0,1,\ldots,N.\label{constraint}
\end{align}
Equation (\ref{constraint}) has a two-fold meaning. First, it allows us to construct another homogeneous $T-Q$ relations and Bethe ansatz equations. Second, one can find a set of vectors which form an invariant subspace and can be used to expand certain eigenstates of the transfer matrix, with the expansion coefficients parameterizable by the Bethe roots.

\begin{rem}
Equation (\ref{constraint}) signifies that a particle has equal transition probabilities to pass through the $M$-particle occupied system from both the left and right boundaries.
\end{rem}

\textit{$T-Q$ relation III: ~~~} Under constraint (\ref{constraint}), using the off-diagonal Bethe ansatz method \cite{WangBook}, another $T-Q$ relation can be constructed
\begin{align}
&\Lambda_3(x)= a_1(x)\,\frac{q^{-\widetilde M}Q_3(qx)}{Q_3(x)}
+ d_1(x)\,\frac{q^{\widetilde M}Q_3(q^{-1}x)}{Q_3(x)},\qquad \widetilde{M}=N-1-M,\label{TQ-3}
\end{align}
where
\begin{align}
Q_3(x)=\prod_{k=1}^{\tM}(\mu_k-x)(\mu_kx-q).\label{Q-3}
\end{align}
The corresponding BAEs are
\begin{align}
\frac{d_1(\mu_j)}{a_1(\mu_j)}=-\prod_{k=1}^{\tM}\frac{(\mu_k-q\mu_j)(\mu_k\mu_j-1)}
{(\mu_k-q^{-1}\mu_j)(\mu_k\mu_j-q^2)},\quad j=1,\ldots,\tM.\label{BAE-3}
\end{align}
The eigenvalue of Markov transition matrix $\mM$ is given by
\begin{align}
E=\frac{(1-q)^2}{q+1}\sum_{k=1}^{\tM}\frac{\mu_k}
{(\mu_k-1)(\mu_k-q)}-(\alpha+\beta+\gamma+\delta).\label{E-3}
\end{align}
\textit{$T-Q$ relation IV: ~~~} Another alternative $T-Q$ relation is
\begin{align}
&\Lambda_4 (x)= a_{2}(x)\frac{q^{-M}Q_{4}(qx)}{Q_{4}(x)}
+ d_{2}(x)\frac{q^MQ_{4}(q^{-1}x)}{Q_{4}(x)},\qquad\label{TQ-4}
\end{align}
where
\begin{align}
Q_4(x)=\prod_{k=1}^{M}(\nu_k-x)(\nu_kx-q).\label{Q-4}
\end{align}
The corresponding BAEs read
\begin{align}
\frac{d_2(\nu_j)}{a_2(\nu_j)}=-\prod_{k=1}^{M}\frac{(\nu_k-q\nu_j)(\nu_k\nu_j-1)}
{(\nu_k-q^{-1}\nu_j)(\nu_k\nu_j-q^2)},\quad j=1,\ldots,M.\label{BAE-4}
\end{align}
The eigenvalue of Markov transition matrix $\mM$ is given by
\begin{align}
E=\frac{(1-q)^2}{q+1}\sum_{k=1}^{M}\frac{\nu_k}
{(\nu_k-1)(\nu_k-q)}.\label{E-4}
\end{align}

 Based on the numerical results obtained from small-scale systems in Tables \ref{Lambda3Lambda4-M1},  \ref{Lambda3Lambda4-M2},\ref{Lambda3Lambda4-M3}, \ref{Lambda3Lambda4-M4}, we conjecture that \textit{
$T-Q$ relation III} in (\ref{TQ-3}) and \textit{
$T-Q$ relation IV} in (\ref{TQ-4}) also consitute the \textit{complete} spectrum of the transfer matrix (the Markov transition matrix). It should be noted that the steady state always corresponds to a specific solution where all the Bethe roots go to infinity, as discussed in detail in Eq. (\ref{string}). 
The functions $\Lambda_3(u)$ and 
$\Lambda_4(u)$ correspond to the subspaces $\mathcal{H}_3$ and $\mathcal{H}_4$, respectively. The left eigenstates belonging to $\mathcal{H}_3$ and the right eigenstates belonging to $\mathcal{H}_4$
can be constructed via the Bethe ansatz approach. 
In this paper, we focus on \textit{$T-Q$ relation IV} in Eq. (\ref{TQ-4}) in order to construct the right steady state. A summary of the invariant subspace and $T-Q$ relation is given in Table \ref{Summary}.

\begin{table}
\centering
\begin{tabular}{|c|c|c|c|c|}
\hline 
Subspace & Left eigenstates & Right eigenstates & Dimension & $T-Q$ relation \\
\hline 
$\mathcal{H}_1$ & -- & Bethe state & $2^N-1$ & I \\
\hline 
$\mathcal{H}_2$ & Bethe state & -- & $1$ & II \\
\hline 
$\mathcal{H}_3$ & Bethe state & -- & $2^N-\sum\limits_{n=0}^{M}\binom{N}{n}$ & III\\
\hline 
$\mathcal{H}_4$ & -- & Bethe state & $\sum\limits_{n=0}^{M}\binom{N}{n}$ & IV \\
\hline
\end{tabular}
\caption{Summary of the invariant subspace, $T-Q$ relation, and the Bethe-type eigenstate.}\label{Summary}
\end{table}

\section{Chiral basis and the coordinate Bethe ansatz}\label{sec;Basis}

\textit{Factorized chiral vectors~~~} Define the following chiral vectors
\begin{align}
\ket{n_1,n_2,\ldots,n_m}=&\bigotimes_{k_1=1}^{n_1}\phi_{k_1}(k_1-1)\bigotimes_{k_2=n_1+1}^{n_2}\phi_{k_2}(k_2-2)\cdots \bigotimes_{k_m=n_{m-1}+1}^{n_m}\phi_{k_m}(k_m-m)\no\\
&\bigotimes_{k_{m+1}=n_m+1}^{N}\phi_{k_{m+1}}(k_{m+1}-m-1),\qquad \phi(x)=\binom{\gamma}{\alpha q^x}.
\label{Basis;M}
\end{align}
The local vector $\phi(x)$ in Eq. (\ref{Basis;M}) satisfies the following relations \cite{PhantomBetheRoots,CCBA}
\begin{align}
&\mM_{n,n+1}\,\phi_n(x)\phi_{n+1}(x+1)=0,\label{div;1}\\
&\mM_{n,n+1}\,\phi_n(x)\phi_{n+1}(x)=\frac{q-1}{2(q+1)}\left(\sigma_n^z-\sigma_{n+1}^z\right)\phi_n(x)\phi_{n+1}(x),\label{div;2}\\
&\mM_1\,\phi_1(x)=\frac{(1-q^{-x})(\alpha  q^x-\gamma)}{2} \,\phi_1 (x)+\frac{(1-q^{-x})(\alpha q^x+\gamma)}{2}  \,\sigma_1^z\,\phi_1 (x),\label{div;left}\\
&\mM_N\,\phi_N(x)=\frac{\left(\alpha  q^x-\gamma \right) \left(\alpha  \beta -\gamma  \delta  q^{-x}\right)}{2 \alpha  \gamma }\phi_N (x)+\frac{\left(\alpha  q^x+\gamma\right) \left(\alpha  \beta -\gamma  \delta  q^{-x}\right)}{2 \alpha  \gamma }\, \sigma_N^z\,\phi_N (x),\label{div;right}\\
&\sigma^z\phi (x)=\frac{q+1}{q-1}\,\phi (x)-\frac{2}{q-1}\,\phi(x+1)=\frac{1+q}{1-q}\,\phi (x)+\frac{2q}{q-1}\,\phi (x-1).\label{sigma;phi}
\end{align}

\textit{Chiral basis of $\mathcal{H}_4$~~~}
Using Eqs. (\ref{div;1}) - (\ref{sigma;phi}) repeatedly, one see that the following vectors
\begin{align}
\ket{\!\underbrace{0,\ldots,0}_{m_0},\underbrace{j_1,j_2,\ldots,j_k}_k,\underbrace{N,\ldots,N}_{m_N}},\quad 0< j_1<j_2\cdots <j_k<N,\quad m_0+k+m_N=M,\quad m_0,k,m_N\geq 0,\label{Basis;M0}
\end{align}
form an invariant subspace of the Hilbert space $\mathcal{H}_4$ under the constraint (\ref{constraint}). It should be noted that the local state on the first site in $\ket{\!\underbrace{0,\ldots,0}_{m_0},\underbrace{j_1,j_2,\ldots,j_k}_k,\underbrace{N,\ldots,N}_{m_N}}$ is $\phi_1(-m_0)$.

\begin{figure}
\begin{tikzpicture}[x=1.0cm,y=1.5cm]
\draw[thick,->] (0,0) -- (6,0) node[anchor=west] {$n$};
\draw[thick,->] (0,0) -- (0,4.0) node[anchor=south] {$z_n$};
\foreach \x in {1,2,3,4,5}
\draw[thick,dashed,gray] (\x,3.5)--(\x,0);
\draw[thick] (0,0.01)--(0,0) node[below=3.6pt]{1};
\draw[thick] (5,0.01)--(5,0) node[below=3.6pt]{$N$};
\draw[thick] (0.01,0)--(0,0) node[left=3.6pt]{$-2$};
\draw[thick] (0.01,0.5)--(0,0.5) node[left=3.6pt]{$-1$};
\draw[thick] (0.01,1.0)--(0,1.0) node[left=3.6pt]{$0$};
\foreach \y in {0.5,1,1.5,2,2.5,3,3.5} \draw[thick,dashed,gray] (5,\y)--(0,\y);
\draw [drop shadow={shadow xshift=0pt,shadow yshift=0pt,fill=gray!50},dotted]
 (0,0)--(5,2.5)--(5,3.5)--(0,1.0)--cycle;
\draw[color=blue,line width=1.0pt] (0,0)--(5,2.5);\coordinate[label=right:$\ket{0,0}$] (a) at (5.2,2.4);
\draw[color=green,line width=1.0pt] (0,1)--(5,3.5);\coordinate[label=right:$\ket{N,N}$] (a) at (5.2,3.5);
\draw[color=purple,line width=1.0pt] (0,0.5)--(5,3);\coordinate[label=right:$\ket{0,N}$] (a) at (5.2,3);
\end{tikzpicture}\qquad
\begin{tikzpicture}[x=1.0cm,y=1.5cm]
\draw[thick,->] (0,0) -- (6,0) node[anchor=west] {$n$};
\draw[thick,->] (0,0) -- (0,4.0) node[anchor=south] {$z_n$};
\foreach \x in {1,2,3,4,5}
\draw[thick,dashed,gray] (\x,3.5)--(\x,0);
\draw[thick] (0,0.01)--(0,0) node[below=3.6pt]{1};
\draw[thick] (1,0.01)--(1,0) node[below=3.6pt]{2};
\draw[thick] (2,0.01)--(2,0) node[below=3.6pt]{3};
\draw[thick] (3,0.01)--(3,0) node[below=3.6pt]{4};
\draw[thick] (4,0.01)--(4,0) node[below=3.6pt]{5};
\draw[thick] (5,0.01)--(5,0) node[below=3.6pt]{$N$};
\draw[thick] (0.01,0)--(0,0) node[left=3.6pt]{$-2$};
\draw[thick] (0.01,0.5)--(0,0.5) node[left=3.6pt]{$-1$};
\draw[thick] (0.01,1.0)--(0,1.0) node[left=3.6pt]{$0$};
\foreach \y in {0.5,1,1.5,2,2.5,3,3.5} \draw[thick,dashed,gray] (5,\y)--(0,\y);
\draw [drop shadow={shadow xshift=0pt,shadow yshift=0pt,fill=gray!50},dotted]
 (0,0)--(5,2.5)--(5,3.5)--(0,1.0)--cycle;
\draw[color=red,line width=1.0pt] (0,1+0) --(1,1.5)--(2,1.5)--(4,2.5)--(5,2.5);\coordinate[label=right:$\ket{2,5}$] (a) at (5.3,2.3);
\draw[color=red,line width=1.0pt,-{To[length=2mm,width=1.5mm]}] (5,2.5)--(5.4,2.3);
\draw[color=black,line width=1.0pt] (0,0.5)--(3,2.0)--(4,2)--(5,2.5);\coordinate[label=right:$\ket{0,4}$] (a) at (5.3,2.6);
\draw[color=black,line width=1.0pt,-{To[length=2mm,width=1.5mm]}] (5,2.5)--(5.4,2.6);
\end{tikzpicture}
\caption{Visualization of the chiral vectors in (\ref{Basis;M0}) with $M=2$. Here $x$-axis is the site number while $y$-axis represents the phase factor $z_{n}$. Each possible trajectory (directed paths) in the shaded region represents a vector in (\ref{Basis;M0}).}\label{Fig;Basis}
\end{figure}
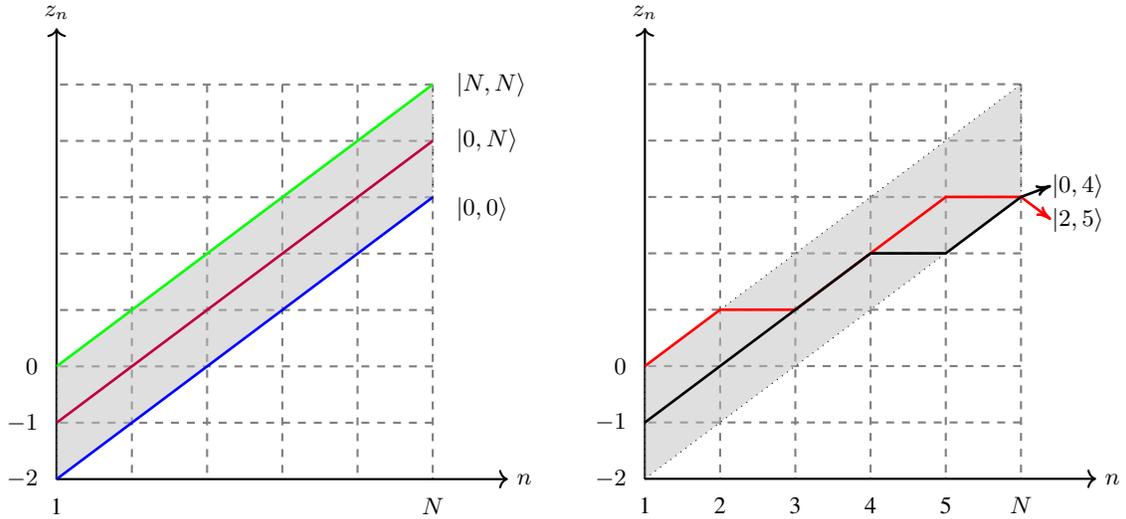

To demonstrate the vectors in (\ref{Basis;M0}) more intuitively, we use a set of phase factor $\{z_1,\ldots,z_n\}$ to represent the factorized state 
\begin{align} 
\bigotimes_{n=1}^N\binom{\gamma}{\alpha q^{z_n}}.
\end{align}
For the chiral vectors in (\ref{Basis;M0}), we can easily get that 
\begin{align}
z_1=0,-1,\ldots,-M,\quad z_{n+1}-z_n=0,1.
\end{align}
The phases of the qubit increment by an amount 1 from site $n$ to site $n+1$ except at the points $n_1,\ldots,n_M$ where kinks occur.
The tensor product state $\ket{n_1,n_2,\ldots,n_M}$ is completely determined by the phase on the first site $z_1$ and the position of the kink and the integer $M$ is the maximum number of kinks. Our chiral vectors with kinks exhibit a structure similar to the Bernoulli product measures described in Refs. \cite{Pigorsch2000, Belitsky2002, Belitsky2013, Schutz2023}. The visualization of our chiral vectors in (\ref{Basis;M0}) is illustrated in Figure \ref{Fig;Basis}. 

It should be noted that only $\sum_{n=0}^{M}\binom{N}{n}$ vectors in the set (\ref{Basis;M0}) are linearly independent, which matches the dimension of the subspace $\mathcal{H}_4$.  For example, the linearly independent basis vectors can be selected as follows
\begin{align}\label{BasisVectors}
\begin{aligned}
&\ket{\!\underbrace{0,\ldots,0}_{m_0},\underbrace{j_1,j_2,\ldots,j_k}_{k}},\quad m_0+k=M,\quad  m_0=0,\ldots,M, \quad 1\leq j_1<j_2\cdots<j_k\leq N,\\
{\rm or},\quad &\ket{\underbrace{j_1,j_2,\ldots,j_k}_k,\underbrace{N,\ldots,N}_{m_N}},\quad m_N+k=M,\quad m_N=0,\ldots,M,\quad 0\leq j_1<j_2\cdots<j_k\leq N-1.
\end{aligned}
\end{align} 
Adding the auxiliary vectors makes the entire set in Eq. (\ref{Basis;M0}) symmetric and leads to explicit and symmetric forms for the expansion coefficients in the Bethe state \cite{CCBA}.

The difference between our chiral vectors (\ref{Basis;M}) and the tensor product states with excitations proposed in Ref. \cite{Simon2009} needs to be clarified. Let us recall the tensor product state in Ref. \cite{Simon2009} with our notation
\begin{align}
\ket{n_1,n_2,\ldots,n_m}_t=&\bigotimes_{k_1=1}^{n_1-1}\phi_{k_1}(k_1-1)\,\varphi_{n_1}(t,n_1-1)\,\bigotimes_{k_2=n_1+1}^{n_2-1}\phi_{k_2}(k_2-2)\,\varphi_{n_2}(t,n_2-2)\,\bigotimes_{k_3=n_2+1}^{n_3-1}\phi_{k_3}(k_3-3)\no\\
&\cdots \bigotimes_{k_m=n_{m-1}+1}^{n_m-1}\phi_{k_m}(k_m-m)\,\varphi_{n_m}(t,n_m-m)\bigotimes_{k_{m+1}=n_m+1}^{N}\phi_{k_{m+1}}(k_{m+1}-m-1),\label{old_basis}\\
\varphi(t,x)=&\,t\phi(x)+(1-t)\phi(x-1),\qquad t\in\mathbb{C}.\label{varphi}
\end{align}
It can be easily verified that $\ket{n_1, n_2, \ldots, n_m}_t$ is always a linear combination of the chiral vectors $\{\ket{n'_1, n'_2, \ldots, n'_{m}}\}$. In particular, when $t=1$, we have
$\ket{n_1,n_2,\ldots,n_m}_t=\ket{n_1,n_2,\ldots,n_m}$. 

We see that the vectors constructed in Refs. \cite{Simon2009,Crampe2010} to expand the Bethe vectors are all linearly independent, similar to those in Eq. (\ref{BasisVectors}). In contrast, auxiliary vectors are added in Eq. (\ref{Basis;M0}) to obtain an enlarged and symmetric set. This significant change of basis results in a different parameterization of the Bethe state. For large $M$, the number of auxiliary vectors is comparable to the number of independent vectors, and it is generally non-trivial to map the basis used in Refs. \cite{Simon2009, Crampe2010} onto our enlarged basis, or vice versa. 

\textit{Chiral coordinate Bethe ansatz~~~} Define the quasi-momentum $\mathrm{p}_j$
\begin{align}
\eE^{\ir\mathrm{p}_j}=\frac{\nu_j-q}{\sqrt{q}(\nu_j -1)},\label{quasi-momentum}
\end{align}
where $\{\nu_1,\ldots,\nu_M\}$ is the solution of BAEs (\ref{BAE-4}). 

One can thus use the chiral basis (\ref{Basis;M0}) and the quasi-momentum $\{\mathrm{p}_j\}$ to construct the Bethe-type eigenstate
\begin{align}
\ket{\Psi(\mathrm{p}_1,\ldots,\mathrm{p}_M)}=&\sum_{n_1,\ldots,n_M}\,\sum_{r_1,\ldots,r_M}\,\sum_{\sigma_1,\ldots,\sigma_M=\pm1}Y_{n_1,\ldots,n_M}A_{\sigma_{r_1},\ldots,\sigma_{r_M}}^{r_1,\ldots,r_M}\no\\
&\times q^{-\frac12\sum_k n_k}\exp\left[\ir \sum_{k=1}^M\sigma_{r_k}n_k\mathrm{p}_{r_k}\right]\ket{n_1,n_2,\ldots,n_M},\label{CCBA}
\end{align}
where $r_1,\ldots,r_M$ is the permutation of $1,\ldots,M$, $Y_{n_1,\ldots,n_M}$ is a certain coefficient independent of $\{\mathrm{p}_j\}$ and the amplitudes $\Big\{A_{\sigma_{r_1},\ldots,\sigma_{r_M}}^{r_1,\ldots,r_M}\Big\}$ are determined by the two-body scattering matrix
$S_{j,k}$ and the boundary reflection matrices $S_L$ and $S_R$ as follows \cite{CCBA}
\begin{align}
\frac{A_{\ldots,\sigma_k,\sigma_j,\ldots}^{\ldots,k,j,\ldots}}{A_{\ldots,\sigma_j,\sigma_k,\ldots}^{\ldots,j,k,\ldots}}=S_{j,k}(\sigma_j\mathrm p_j,\sigma_k\mathrm p_k),\quad \frac{A_{-1,\ldots}^{j,\ldots}}{A_{+1,\ldots}^{j,\ldots}}=S_L(\mathrm p_j),\quad \frac{A_{\ldots,-1}^{\ldots,k}}{A_{\ldots,+1}^{\ldots,k}}=\eE^{2\ir N\mathrm{p}_k}S_R(\mathrm p_k).
\end{align}
This is the logic of the chiral coordinate Bethe ansatz method in Refs. \cite{PhantomBetheRoots,CCBA}. More details about the construction of Bethe state (\ref{CCBA}) are demonstrated in Ref. \cite{CCBA}. In this paper, we only focus on the right steady state. 

\textit{Specific Bethe roots for the steady state~~~} The steady state $\ket{\Phi}$ corresponds to a special solution of BAEs (\ref{BAE-4}), i.e., all the Bethe roots $\{\nu_1,\ldots,\nu_M\}$ go to infinity and form the following string 
\begin{align}
\nu_j=\eE^{\frac{2\ir\pi (j-k)}{M}}\nu_k,\quad \nu_l\to +\infty.\label{string}
\end{align}
\begin{proof}
 Substituting the string solution in (\ref{string}) into Eq. (\ref{BAE-4}), one can verify that the left side reads 
\begin{align}
{\rm LHS}=\frac{d_2(\nu_j)}{a_2(\nu_j)}={abcd q^{N-1}}=q^{M}.
\end{align}
The right side of (\ref{BAE-4}) equals to 
\begin{align}
{\rm RHS}&=-\prod_{k=1}^{M}\underbrace{\frac{(\nu_k-q\nu_j)}
{(q\nu_k-\nu_j)}}_{ S_{j,k}(\mathrm{p}_j,\mathrm{p}_k)}\underbrace{\frac{(\nu_k\nu_j-1)}
{(q^{-1}\nu_k\nu_j-q)}}_{S_{j,k}(\mathrm{p}_j,-\mathrm{p}_k)}=-q^M\prod_{k=1}^{M}\underbrace{\frac{(1-q\nu_j/\nu_k)}
{(q-\nu_j/\nu_k)}}_{S_{j,k}(\mathrm{p}_j,\mathrm{p}_k)}\no\\[2pt]
&=-q^{M}\prod_{k=1}^M\frac{(1-q r^{j-k})}{(q-r^{j-k})}=q^{M}\prod_{k=0}^{M}\frac{(1-q r^{k})}{(q-r^{-k})}\no\\
&=q^{M}\prod_{k=0}^{M}\frac{(1-q r^{k})}{r^k(qr^k-1)}=q^{M}\prod_{k=0}^M r^{-k+\frac{M}{2}}=q^M,\label{String;Proof}
\end{align}
where $r=\eE^{\frac{2\ir\pi}{M}},\,r^M=1$. The string solution in (\ref{string}) is thus proved.
\end{proof}

For the steady state, the quasi-momentum tends to be nearly identical to each other
\begin{align}
\eE^{\ir\mathrm{p}_j}\to\frac{1}{\sqrt{q}}+\epsilon_j,\quad j=1,\ldots,M,
\end{align} 
where $\{\epsilon_j\}$ are infinitesimal numbers.
In this case, since $S_{j,k}(\mathrm{p}_j,\mathrm{p}_k)$ is now a constant not equal to $-1$ (see Eq. (\ref{String;Proof})), we no longer need to consider the bulk scattering $\pm\mathrm{p}_j,\pm\mathrm{p}_k\to \pm\mathrm{p}_k,\pm\mathrm{p}_j$ anymore, only the reflection $\pm\mathrm{p}_j\to \mp\mathrm{p}_j$ and the bulk scattering $\pm\mathrm{p}_j,\mp\mathrm{p}_k\to \mp\mathrm{p}_k,\pm\mathrm{p}_j$ contribute non-trivially. The wave-function (the expansion coefficient) in the Bethe-type steady state therefore has much simpler expansion than the generic one in (\ref{CCBA}). 
The explicit expression of the right steady state is shown in Section \ref{sec;SteadyState}.

\textit{Chiral basis of $\mathcal{H}_1$~~~} Define another factorized chiral vectors
\begin{align}
\kket{l_1,l_2,\ldots,l_{m}}=&\bigotimes_{k_1=1}^{l_1}\tilde\phi_{k_1}(k_1-1)\bigotimes_{k_2=l_1+1}^{l_2}\tilde\phi_{k_2}(k_2-2)\cdots \bigotimes_{k_{m}=l_{m-1}+1}^{l_{m}}\tilde\phi_{k_{m}}(k_{m}-m)\no\\
&\bigotimes_{k_{m+1}=l_{m}+1}^{N}\tilde\phi_{k_{m+1}}(k_{m+1}-m-1),\qquad \tilde{\phi}(x)=\phi(x+x_0),\qquad q^{x_0}=-\frac{\gamma}{\alpha}.
\label{Basis;2}
\end{align}
Using Eqs. (\ref{div;1}) - (\ref{sigma;phi}) repeatedly, one see that the following vectors
\begin{align}
\kket{\!\underbrace{0,\ldots,0}_{m_0},\underbrace{j_1,j_2,\ldots,j_k}_k,\underbrace{N,\ldots,N}_{m_N}},\quad 0< j_1<j_2\cdots <j_k<N,\quad m_0+k+m_N=N-1,\quad m_0,k,m_N\geq 0,\label{Basis;3}
\end{align}
always form an invariant subspace of the Hilbert space, i.e., $\mathcal{H}_1$. The vectors in (\ref{Basis;M0}) and (\ref{Basis;3}) have a similar structure, therefore we can construct the right Bethe state belonging to $\mathcal{H}_1$ using the same technique shown in (\ref{CCBA}).

\section{Bethe ansatz for the steady state}\label{sec;SteadyState}
\subsection{\texorpdfstring{$M=0$ case: Factorized steady state}{M=0 case: Factorized steady state}}

When $M=0$, we can construct a factorized right steady state 
\begin{align}
\ket{\Phi_0}=\bigotimes_{n=1}^N\binom{\gamma}{\alpha q^{n-1}}.
\end{align}
In this case, although the Markov transition matrix $\mathcal{M}$ remains non-Hermitian, the XXZ Hamiltonian (\ref{H_XXZ}) is Hermitian. Therefore, one can use the Bethe ansatz to construct \textit{all} the eigenstates of $\mathcal{M}$.

Introduce the current operator at the left boundary   
\begin{align}
\hat{j}=\left(
\begin{array}{cc}
0 & -\gamma \\
\alpha & 0
\end{array}\right).\label{J}
\end{align}
It is straightforward to prove that the current in the steady state is zero
\begin{align}
\langle \hat j\rangle =\frac{\bra{\Phi}\hat{j}\ket{\Phi_0}}{\braket{\Phi}{\Phi_0}}=0.
\end{align}
The density profile now has a very simple form 
\begin{align}
\langle \hat{n}_k\rangle=\frac{\alpha q^{k-1}}{\gamma+\alpha q^{k-1}},\qquad \hat{n}=\left(\begin{array}{cc}
0 & 0\\
0 & 1
\end{array}\right),
\end{align} 
and demonstrates a skin-like effect as follows
\begin{align}
\begin{aligned}
\langle \hat{n}_1\rangle<\langle \hat{n}_2\rangle<\cdots<\langle \hat{n}_N\rangle,\quad \mbox{when} \quad q>1,\\
\langle \hat{n}_1\rangle>\langle \hat{n}_2\rangle>\cdots>\langle \hat{n}_N\rangle,\quad \mbox{when} \quad q<1.
\end{aligned}
\end{align}

\subsection{\texorpdfstring{$M=1$ case}{M=1 case}}
When $M=1$, the set (\ref{Basis;M0}) contains $N+1$ vectors. With the help of Eqs. (\ref{div;1})-(\ref{sigma;phi}), we can derive that 
\begin{align}
\begin{aligned}
&\mM\ket{n}=-\ket{n}+\frac{q }{q+1}\ket{n-1}+\frac{1}{q+1}\ket{n+1},\quad n=1,\ldots,N-1,\\
&\mM\ket{0}=-(\alpha+\gamma)\ket{0}+\frac{\alpha + q\gamma}{q}\,\ket{1},\\
&\mM\ket{N}=-(\beta+\delta)\ket{N}+(\beta + q\delta)\ket{N-1}.
\end{aligned}
\end{align}
Suppose that the steady state can be expanded by $\{\ket{0},\ket{1},\ldots,\ket{N}\}$ as 
\begin{align}
\ket{\Phi_1}=\sum_{n=0}^{N}\chi_n \ket{n}.
\end{align} 
The eigen equation $\mathcal{M}\ket{\Phi_1}=0$ gives rise to the following recursive equations for the expansion coefficients $\{\chi_0,\ldots,\chi_N\}$ 
\begin{align}\label{eq;recusive}
\begin{aligned}
&\chi_n=\frac{q}{q+1}\chi_{n+1}+\frac{1}{q+1}\chi_{n-1},\quad n=2,\ldots,N-2,\\
&\chi_1=\frac{q}{q+1}\chi_{2}+\frac{\alpha + q\gamma}{q}\chi_0,\\
&\chi_{N-1}=(\beta + q\delta)\chi_N+\frac{1}{q+1}\chi_{N-2},\\
&(\alpha+\gamma)\chi_0=\frac{q}{q+1}\chi_1,\\
&(\beta+\delta)\chi_N=\frac{1}{q+1}\chi_{N-1}.
\end{aligned}
\end{align} 
The aforementioned equations (\ref{eq;recusive}) allow us to propose the following ansatz
\begin{align}
&\chi_n=f_n,\qquad n=1,\ldots,N-1,\no\\
&\chi_0=g_{\rm l}(1)f_0,\qquad \chi_N=g_{\rm r}(1)f_N,\label{ansatz;M1}
\end{align}
where 
\begin{align}
&f_n=\gamma q^{-n}+\alpha  q^{-1},\\
&g_{\rm l}(n)=\frac{q-1}{(q+1) \left(q^n-1\right) \left(\gamma +\alpha  q^{-n}\right)}-\frac{\left(q^{n-1}-1\right) \left(\gamma +\alpha  q^{-n+1}\right)}{\left(q^n-1\right) \left(\gamma +\alpha  q^{-n}\right)},\label{gl}\\
&g_{\rm r}(n)=\frac{q-1}{(q+1) \left(q^n-1\right) \left(\delta +\beta  q^{-n}\right)}-\frac{\left(q^{n-1}-1\right) \left(\delta +\beta  q^{-n+1}\right)}{\left(q^n-1\right) \left(\delta +\beta  q^{-n}\right)}.\label{gr}
\end{align}
By substituting the ansatz (\ref{ansatz;M1}) into Eq. (\ref{eq;recusive}), our ansatz can be easily proved.

Before calculating the current and density profile, we introduce the following identities 
\begin{align}
&\braket{\Phi}{n}=(\gamma+\alpha q^{n-1})\prod_{k=1}^{N-1}(\gamma+\alpha q^{k-1}),\label{eq;1}\\
&\bra{\Phi}\hat{j}\ket{n}=\delta_{n,0}\,\alpha\gamma(1-q^{-1})\prod_{k=1}^{N-1}(\gamma+\alpha q^{k-1}).\label{eq;2}
\end{align}
Using Eqs. (\ref{eq;1}) and (\ref{eq;2}), the expectation value of the current can be obtained 
\begin{align}
\langle \hat j\rangle =\frac{\bra{\Phi}\hat{j}\ket{\Phi_1}}{\braket{\Phi}{\Phi_1}}=-\frac{(q-1)^2}{2 (N-1) \left(1-q^2\right)+(q+1) \left(\frac{\alpha }{\gamma }+\frac{\beta }{\delta }-\frac{\gamma  q}{\alpha }-\frac{\delta  q}{\beta }\right)+(1-q) \left(\frac{1}{\gamma }+\frac{1}{\delta }+\frac{q}{\alpha }+\frac{q}{\beta }\right)}.
\end{align}
One can also derive the density profile
\begin{align}
\langle \hat{n}_k\rangle&=1+(q^2-1) \alpha^{-1} \left[\frac{\alpha  \gamma\beta^{-1}\delta^{-1}  \left(\beta ^2+2 \beta  \delta  (q-1) (k-N)-\delta ^2 q\right)+\alpha ^2 q^{k-1}-\gamma ^2 q^{2-k}}{(1-q) \left(\gamma +\alpha  q^{k-1}\right)}\right.\no\\
&\quad\left.+\frac{\alpha ^2 q^{k-1}-\gamma ^2 q^{2-k}-\alpha ^2+2 \alpha  \gamma  (k-1) (q-1)+\gamma ^2 q}{(q-1) \left(\gamma +\alpha  q^{k-2}\right)}+\frac{\alpha +\gamma  q}{(q+1) \left(\gamma +\alpha  q^{k-2}\right)}+\frac{\alpha  \gamma  (\beta +\delta  q)}{\beta  \delta  (q+1) \left(\gamma +\alpha  q^{k-1}\right)}\right]\no\\
&\quad \times \left[2 (N-1) (1-q^2)+(q+1) \left(\frac{\alpha }{\gamma }+\frac{\beta }{\delta }-\frac{\gamma  q}{\alpha }-\frac{\delta  q}{\beta }\right)+(1-q) \left(\frac{1}{\gamma }+\frac{1}{\delta }+\frac{q}{\alpha }+\frac{q}{\beta }\right)\right]^{-1}.
\end{align}
\textit{Large $N$ case~~~}
For a large-scale system,  once finite values for $\alpha$, $\gamma$ and $q$ are selected, we observe that $\beta \to 0$ when $q > 1$ and $\delta \to 0$ when $q < 1$. This leads to a small current
\begin{align}
&\langle\hat{j}\rangle\approx\frac{\gamma  \delta  (q-1)^2 q^{1-N}}{\alpha  (\delta +\delta  q+q-1)},\quad \mbox{when}\,\, q>1,\,\,\beta\to 0,\\
&\langle\hat{j}\rangle\approx-\frac{\alpha  \beta  (q-1)^2 q^{N-2}}{\gamma  (\beta +\beta q -q+1)},\quad \mbox{when}\,\, q<1,\,\,\delta\to 0.
\end{align}

\subsection{\texorpdfstring{Arbitrary $M$ case}{Arbitrary M case}}

By following the same procedure, namely, deriving the recursive relation for the expansion coefficients, proposing a proper ansatz, and verifying its correctness, the Bethe-type steady state can also be constructed for cases where $M>1$. We present the expression for the steady state directly.

Under the constraint (\ref{constraint}), the steady state of the system $\ket{\Phi_M}$ can be expanded as a linear combination of the vectors in (\ref{Basis;M0}) as follows
\begin{align}
&\ket{\Phi_M}=\sum_{m_0,m_N}\sum_{j_1,\ldots,j_k}\kappa_{0,\ldots,0,j_1,\ldots,j_k,N,\ldots,N}\,f_{0,\ldots,0,j_1,\ldots,j_k,N,\ldots,N}\ket{\underbrace{0,\ldots,0}_{m_0},\underbrace{j_1,j_2,\ldots,j_k}_k,\underbrace{N,\ldots,N}_{m_N}},\label{SS;M}\\
&0< j_1<j_2\cdots <j_k<N,\quad m_0+k+m_N=M,\quad m_0,k,m_N\geq 0.\no
\end{align}
The coefficient $ f_{0,\ldots,0,j_1,\ldots,j_k,N,\ldots,N}\equiv f_{l_1,\ldots,l_M}$ has a factorized form 
\begin{align}
f_{l_1,\ldots,l_M}=\prod_{s=1}^M(\gamma q^{-l_s}+\alpha  q^{-s}).
\end{align}
The prefactor $\kappa$ in Eq. (\ref{SS;M}) depends only on the integers $m_0$ and $m_N$, and is given by the following equation
\begin{align}
\kappa_{0,\ldots,0,j_1,\ldots,j_k,N,\ldots,N}=\prod_{s=1}^{m_0}g_{\rm l}(s)\prod_{s'=1}^{m_N}g_{\rm r}(s'),\quad g_{\rm l,r}(0)\equiv 1,
\end{align}
where $g_{\rm l,r}(m)$ are defined in Eqs. (\ref{gl}) and (\ref{gr}) and possess the following symmetry 
\begin{align}
g_{\rm l}(m)|_{\alpha\to\beta,\gamma\to\delta}=g_{\rm r}(m).
\end{align}

\textit{Current in the $M=2$ case~~~}
When $M\geq 2$, the expression for the current and the density profile become extremely complicated. For the case where $M=2$, after tedious calculation, we derive the expression of current as follows $$\langle \hat{j}\rangle=\frac{\mathfrak{a}_1}{\mathfrak{a}_2},$$ where 
\begin{align}
\mathfrak{a}_1=&\,\alpha  \gamma  \left(q^2-1\right) \left[(q+1)\, q^3 \sum _{m=1}^{N-1} q^{-m} \left(\gamma +\alpha  q^{m-2}\right)^2+q\alpha  \gamma \beta^{-1}\delta^{-1} (\beta +\delta  q)\right.\no\\
&\left.-\left(\alpha +\gamma  q^2+\alpha  q+\gamma  q-q\right) (\alpha +\gamma  q)\right],\\
\mathfrak{a}_2=\,&q^4(q+1)^3\sum _{n=1}^{N-2} \sum _{m=n+1}^{N-1} q^{-m-n} \left(\gamma +\alpha  q^{m-2}\right)^2 \left(\gamma +\alpha  q^{n-1}\right)^2\no\\
&+q^3(q+1)^2\left(q\gamma +\alpha\right) \sum _{m=1}^{N-1} q^{-m} \left(\gamma +\alpha  q^{m-2}\right)^2\no\\
&+q^2(q+1)^2\alpha\gamma  \beta^{-1}\delta^{-1}(\beta +\delta  q) \sum _{n=1}^{N-1} q^{-n} \left(\gamma +\alpha  q^{n-1}\right)^2\no\\
&-\alpha ^2 \gamma ^2 \beta^{-2}\delta^{-2} \left(\beta +\delta  q^2+\beta  q+\delta  q-q\right) (\beta +\delta  q) \left(\beta +\delta  q^2\right)\no\\
&-\left(\alpha +\gamma  q^2+\alpha  q+\gamma  q-q\right) (\alpha +\gamma  q) \left(\alpha +\gamma  q^2\right)\no\\
&+q(q+1)\alpha  \gamma\beta^{-1}\delta^{-1}(\alpha +\gamma  q) (\beta +\delta  q).
\end{align} 
Similar to the $M=1$ case, for large $N$, the current $\langle\hat{j}\rangle$ remains small
\begin{align}
&\langle\hat{j}\rangle\approx\frac{\gamma  \delta  (q-1)^2 (q+1) q^{1-N}}{\alpha  (\delta +\delta  q+q-1)},\quad \mbox{when}\,\, q>1,\,\,\beta\to 0,\\
&\langle\hat{j}\rangle\approx-\frac{\alpha  \beta  (q-1)^2 (q+1) q^{N-3}}{\gamma  (\beta +\beta  q-q+1)},\quad \mbox{when}\,\, q<1,\,\,\delta\to 0.
\end{align}

\section{Generic case}\label{sec;GenericASEP}

For the ASEP with generic open boundaries, our analytical approach fails. To overcome this, we can employ some numerical techniques to study the properties of the steady state more comprehensively.

\begin{figure}[htbp]
\centering
\includegraphics[width=0.45\textwidth]{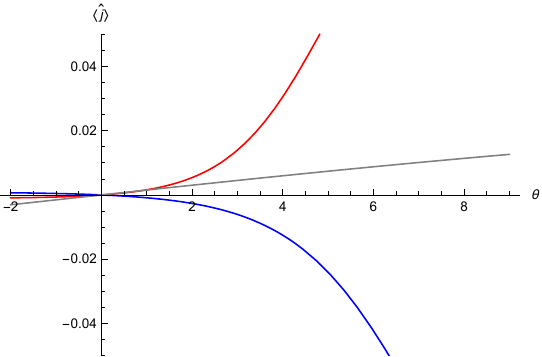}~~~~~~\includegraphics[width=0.45\textwidth]{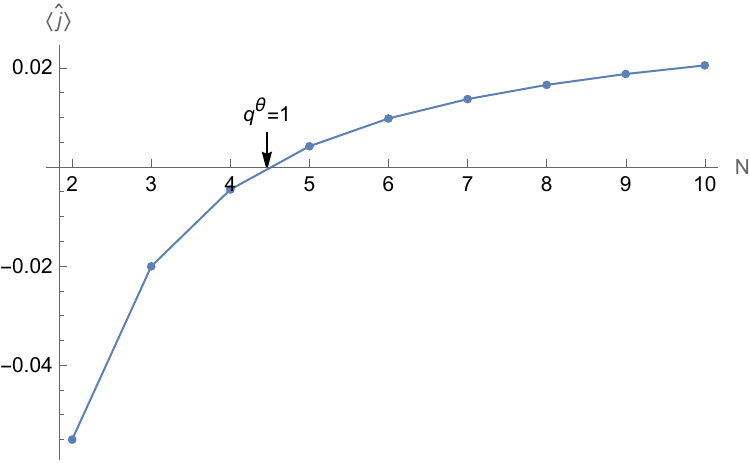}
\caption{Left Panel: The curves of $\langle \hat j\rangle$ versus $\theta$ with $N=8,\alpha=1.30,\beta=0.46, \gamma=2.11$. The blue, red and gray lines correspond to $q=0.5$, $q=2.5$ and $q=1.1$ respectively. Right Panel: The curves of $\langle \hat j\rangle$ versus $N$ with $\alpha=1.30,\beta=0.40,\gamma=2.10,\delta=0.70$ and $q=1.35$. We observe an interesting phenomenon of current reversal as the system size $N$ increases \cite{Bryc2019}.}\label{Fig2}
\end{figure}

Let us define a parameter $\theta$ 
\begin{align}
q^\theta=\frac{\alpha\beta q^{N-1}}{\gamma\delta},
\end{align}
which relates to the ratio of the transition probabilities for a particle to pass through the full-empty system from the left and right boundaries. The current in the steady state $\langle \hat j\rangle$ has the same sign as $q^\theta-1$, as illustrated in Figure \ref{Fig2}.

\begin{figure}[htbp]
\centering
\includegraphics[width=0.45\textwidth]{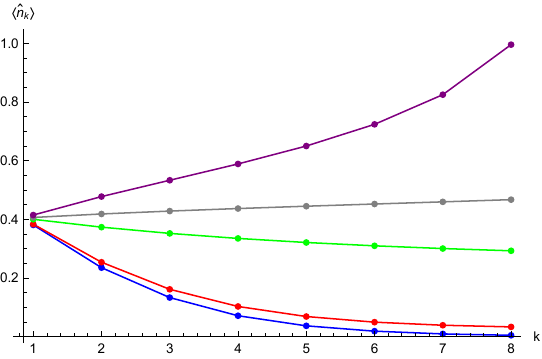}~~~~~~
\includegraphics[width=0.45\textwidth]{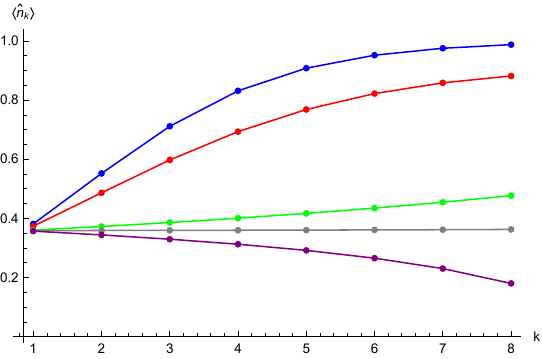}
\caption{The curves of $\langle \hat{n}_k\rangle$ versus $k$. Here we consider the case where $N=8, \alpha=1.30,\beta=0.46, \gamma=2.11$. The blue, red, green, gray and purple lines correspond to the cases $\theta=0$, $\theta=3.5$, $\theta=N-1$, $\theta=N$ and $\theta=2N$ respectively. The left and right panels correspond to $q=0.50$ and $q=2.00$ respectively.}\label{Fig3}
\end{figure}

For the density profile in the steady state, we observe that when $\theta\ll N $ and $\theta\gg N$, it exhibits different skin-like effects. However, when $\theta\sim N$, the skin-like effect will disappear. See Figure \ref{Fig3} for an example.

\begin{figure}[htbp]
\centering
\includegraphics[width=0.6\textwidth]{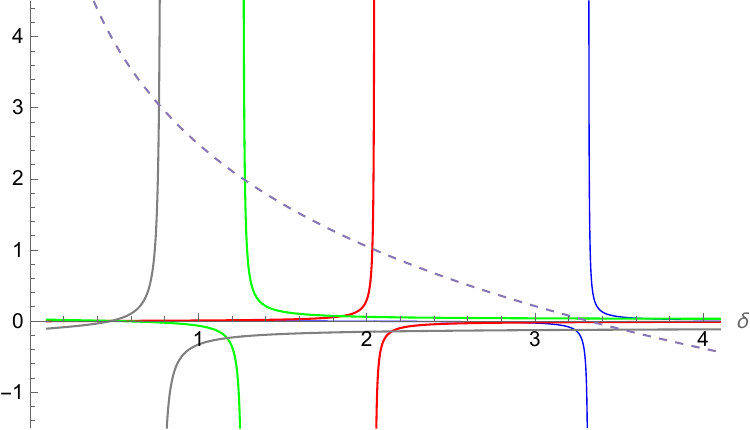}
\caption{The curves of $\{\omega_0/\omega_N,\ldots,\omega_{N-1}/\omega_N\}$ versus $\delta$ with $N=4$, $\alpha=0.51 ,\,\beta=1.27 ,\,\gamma=0.83$ and $q=1.62$.  The blue, red, green and gray lines represent $\omega_0/\omega_4$, $\omega_1/\omega_4$, $\omega_2/\omega_4$ and $\omega_3/\omega_4$ respectively. The dashed line represents the curve of $\theta$ versus $\delta$.}\label{Fig1}
\end{figure}

Numerical results indicate that the steady state $\ket{\Phi}$ can be expanded as follows  
\begin{align}
\ket{\Phi}=\sum_{k=0}^N\omega_k\ket{\Phi_k},\quad \omega_k=\varepsilon_k\prod_{\substack{m=0\\m\neq k}}^N(\alpha\beta q^{N-1-m}-\gamma\delta),
\end{align}
where $\ket{\Phi_k}$, given by Eq. (\ref{SS;M}), is the steady state of the model under the constraint $\alpha\beta q^{N-1-k}=\gamma\delta$, and $\{\varepsilon_k\}$ are some very complicated coefficients. The curves in Figure \ref{Fig1} clearly demonstrate that the state $\ket{\Phi}$ degenerates into $\ket{\Phi_\theta}$ when $\theta$ is an integer and $0\leq \theta\leq N$.

\section*{Discussion}

Under the constraint (\ref{constraint}), we derive the right steady state $\ket{\Phi_k}$. An interesting result is that $\ket{\Phi_k}$ can still be expressed as a Bethe state with Bethe roots forming a special string (\ref{string}), even though the system is non-Hermitian. The key aspect of our approach is the construction of a set of chiral vectors in (\ref{Basis;M0}) on which the Markov transition matrix acts simply. It should be noted that the right eigenstates belonging to $\mathcal{H}_{1}$ and the left eigenstates belonging to $\mathcal{H}_{2,3}$ can be also constructed using the coordinate Bethe ansatz methods, as addressed in Refs. \cite{Crampe2010,CCBA}.

Further clarification is needed to highlight the differences between the chiral coordinate Bethe ansatz used in this paper and the coordinate Bethe ansatz proposed in Refs. \cite{Simon2009, Crampe2010}. As explained in Section \ref{sec;Basis}, the bases in these two approaches are different, leading to distinct parameterizations of the Bethe state. Although the product state with excitations $\ket{n_1,n_2,\ldots,n_m}_t$ in Eq. (\ref{old_basis}) degenerates into our chiral vector when $t=1$, certain functions in the Bethe state constructed in Ref. \cite{Simon2009} will diverge in this case. Therefore, one must take a specific limit, which may be non-trivial. To sum up, these two approaches follow the same logic but are presented differently, and we cannot easily recover one approach from the other.
We prefer the chiral basis in this paper because it leads to a more symmetric expression for the Bethe state. An advantage of the basis vectors in Refs. \cite{Simon2009, Crampe2010} is that they have a well-defined limit when $q\to1 $ and $t\neq 0,1$, whereas our chiral vectors do not exhibit this property.

The right steady state of the ASEP can also be constructed using the matrix product ansatz method \cite{Derrida1993, Blythe2007}. In this approach two operators $\bf D$ and $\bf E$ are introduced, and they satisfy the following relations
\begin{align*}
&q{\bf DE}-\bf{ED} = \bf{D} + \bf{E},\\
&\bra{W}(\alpha {\bf E}-\gamma {\bf{D}})(q+1) = \bra{W},\\
&(q+1)(\beta {\bf D}-\delta {\bf E})\ket{V} = \ket{V},
\end{align*}
where $\ket{V}$ and $\bra{W}$ are infinite dimensional vectors.
Let us introduce the canonical basis $\langle\tau_1,\tau_2, \ldots,\tau_N|$, which is indexed by the occupation values $\tau_i\in\{0,1\}$ on site $i$, where $\tau_i=0$ and $\tau_i=1$ represent a hole and a particle, respectively. 
The stationary probabilities are then computed as
\begin{align*}
P(\tau_1,\ldots,\tau_N)&=\bra{\tau_1,\tau_2,\ldots,\tau_N}\Phi\rangle\\
&=\frac{\bra{W}\prod_{j=1}^N(\tau_j{\bf D} +(1-\tau_j){\bf E})\ket{V}}{\bra{W}({\bf D}+{\bf E})^N\ket{V}}.
\end{align*}
We see that the constraint in Eq. (\ref{constraint}) is a singular point for the matrix product ansatz approach \cite{Essler1996, Bryc2019} and it has been demonstrated that finite-dimensional representations can exist at these specific points \cite{Essler1996, Mallick1997}.

For the ASEP with generic open boundaries, the steady state cannot be expanded using our chiral basis in the same manner. Upon completing this work, we realized that the Bethe-type steady state can be recovered based on the inhomogeneous $T-Q$ relation and the corresponding Bethe ansatz equations, as addressed in Ref. \cite{Zhang2014,WangBook}. The generators and the reference state in the Bethe state can be obtained respectively by employing two sets of gauge transformations. Additionally, the Bethe roots corresponding to the steady state need to be determined. These results will be presented in separate publications.

In the limit $q\to 1$, the open ASEP described in Eq. (\ref{MarkovMatrix}) degenerates into the symmetric simple exclusion process (SSEP) with open boundaries, which can be mapped to an XXX chain with triangular boundary conditions. The algebraic Bethe ansatz approach has been developed to obtain the exact spectrum and eigenstates of the SSEP with open boundaries, as well as the corresponding open XXX model \cite{Belliard2013, Antonio2014, Crampe2014}. It is worth noting that the right steady state of the SSEP with generic open boundary conditions can be expressed as a conventional Bethe state where the number of Bethe roots equals the lattice site number $N$, or equivalently, as the representation in Ref. \cite{Frassek2020}. In the limit $q\to1$, the constraint (\ref{constraint}) studied in this paper reduces to $\alpha\beta =\gamma\delta$, and our chiral vectors become indistinguishable. This leads to a factorized right steady state of the form $\otimes_{n=1}^N \binom{\gamma}{\alpha}$, and the system now is in equilibrium \cite{Frassek2020} with $\langle\hat{j}\rangle=0,\,\, \langle\hat{n}_k\rangle=\frac{\alpha}{\alpha+\gamma}$.

We see that the integer $M$ in Eq. (\ref{constraint}) serves as a conserved charge and equals to the maximum number of the kink. It suggests the presence of a hidden symmetry under Eq. (\ref{constraint}), which will be a focus of our future research. Another intriguing question is the generalization of our approach to multi-species asymmetric simple exclusion processes (m-ASEP) with open boundaries. It is known that homogeneous $T-Q$ relations also exist in m-ASEP \cite{Zhang2019}. A feasible direction for exploration is to identify a constraint analogous to Eq. (\ref{constraint}) in m-ASEP.

\section*{Acknowledgments}

X. Z. acknowledges financial support from the National Natural Science Foundation of China (No. 12204519). F.-K. Wen acknowledges financial support from the National Natural Science Foundation of China (No. 12465001). X. Z. thanks Y. Wang, J. Cao, W.-L. Yang and V. Popkov for valuable discussions. 


\appendix

\section{Numeric solutions of the Bethe ansatz equations}

To improve the readability of the main text, we place the corresponding numerical results in the appendix.

\begin{table}[htbp]
  \centering
\begin{tabular}{|c|c|c|c|c|c|} 
\hline\hline $\lambda_1$ & $\lambda_2$ & $\lambda_3$  & $E$   \\ \hline 
$0.9702-0.9849\ir$  &  $0.0166+0.1513\ir$  &  $0.3762+0.4221\ir$ & $-3.2590+0.5999\ir$  \\ 
 $0.3762-0.4221\ir$  &  $0.9702+0.9849\ir$  &  $0.0717+0.6529\ir$ & $-3.2590-0.5999\ir$  \\ 
 $2.3555-0.9589\ir$  &  $2.2560-0.0000\ir$  &  $2.3555+0.9589\ir$ & $-0.2634$  \\ 
 $1.7812-0.8533\ir$  &  $0.1683-0.6900\ir$  &  $0.3323+0.6834\ir$ & $-2.4104+0.3628\ir$  \\ 
 $0.3323-0.6834\ir$  &  $0.1683+0.6900\ir$  &  $1.7812+0.8533\ir$ & $-2.4104-0.3628\ir$  \\ 
 $-0.2418-0.9907\ir$  &  $2.2672-0.4367\ir$  &  $1.5293-1.6967\ir$ & $-1.3430+0.8503\ir$  \\ 
 $-0.2418+0.9907\ir$  &  $2.2672+0.4367\ir$  &  $1.5293+1.6967\ir$ & $-1.3430-0.8503\ir$  \\ 
 $0.2413+0.6866\ir$  &  $0.2413-0.6866\ir$  &  $2.0174-0.0000\ir$ & $-2.0913$  \\ 
 $0.0314-1.2867\ir$  &  $2.2162-0.1875\ir$  &  $2.2462+0.5359\ir$ & $-0.7852+0.2111\ir$  \\ 
 $2.2162+0.1875\ir$  &  $0.0314+1.2867\ir$  &  $2.2462-0.5359\ir$ & $-0.7852-0.2111\ir$  \\  
 $2.2388+0.3399\ir$  &  $1.7052-1.5918\ir$  &  $-0.1620-1.0905\ir$ & $-1.2358+0.5220\ir$  \\ 
 $2.2388-0.3399\ir$  &  $1.7052+1.5918\ir$  &  $-0.1620+1.0905\ir$ & $-1.2358-0.5220\ir$  \\  
 $0.2499-1.3356\ir$  &  $2.0462-1.1818\ir$  &  $0.9432+1.5458\ir$ & $-1.6171+0.2869\ir$  \\ 
 $2.0462+1.1818\ir$  &  $0.9432-1.5458\ir$  &  $0.2499+1.3356\ir$ & $-1.6171-0.2869\ir$  \\ 
 $0.5919-1.4077\ir$  &  $0.5919+1.4077\ir$  &  $2.1651+0.0000\ir$ & $-1.3044$  \\ 
 --&-- &-- & ~~~0.0000 \\
 \hline\hline\end{tabular}
  \caption{Solutions of BAEs (\ref{BAEs-GC}) for $N=4, q=0.1, \alpha=0.23, \beta=0.32, \gamma=0.47, \delta=0.6$ ($a=1.5466, b=1.3055,
c=-0.3164, d=-0.4085$). The eigenvalues $E$ calculated from Eq.  (\ref{E-lambda1}) are the same as those obtained from the exact diagonalization of the Markov transition matrix $\mathcal{M}$.}\label{Lambda01-com}
\end{table}

\begin{table}[htbp]
  \centering
\begin{tabular}{|c|c|c|c|c|c|} 
\hline\hline $\mu_1$ & $\mu_2$ & $\nu_1$ &   $E$ \\ \hline 
$0.6968+0.1206\ir$  &  $0.5683-0.4208\ir$ & -- & $-2.8518$   \\  
 $3.9600+0.0000\ir$  &  $0.6991-0.1064\ir$ & -- & $-2.3998$   \\ 
 $0.3414+0.0000\ir$  &  $0.7039+0.0669\ir$ & -- & $-2.0923$   \\ 
 $3.7661+0.0000\ir$  &  $0.6575+0.2602\ir$ & -- & $-1.6596$   \\ 
 $0.6739-0.2140\ir$  &  $1.6342+0.0000\ir$ & -- & $-1.5453$   \\  
 $0.2671-0.0215\ir$  &  $0.2671+0.0215\ir$ & -- & $-0.3180$   \\ 
 $0.6623-0.9669\ir$  &  $0.2411-0.3520\ir$ & -- & $-1.0991$   \\ 
 $0.3507+0.6140\ir$  &  $2.7996+0.0000\ir$ & -- & $-0.9242$   \\ 
 $5.8061-0.0000\ir$  &  $1.8539-0.0000\ir$ & -- & $-0.5230$   \\  
 $1.7541+0.0000\ir$  &  $0.4337-0.5585\ir$ & -- & $-0.7826$   \\ 
 $0.0778-0.0000\ir$  &  $2.1292+0.0000\ir$ & -- & $-0.6020$   \\  
 --&--& Solution (\ref{string}) & ~~~$0.0000$  \\ 
 --&--& $0.7026+0.0797\ir$ & $-1.7579$  \\ 
 --&--& $0.1795+0.6839\ir$ & $-0.1461$  \\ 
 --&--& $0.6843-0.1781\ir$ & $-1.2688$  \\ 
 --&--& $0.6229+0.3347\ir$ & $-0.6554$  \\ 
 \hline\hline\end{tabular} 
  \caption{Solutions of BAEs (\ref{BAE-3}) and (\ref{BAE-4}) for $N=4, \widetilde{M}=2, M=1, q=0.5, \alpha=0.23, \beta=0.32, \gamma=0.17$ ($a=2.7974, b=5.5673,c=-0.4836,d=-0.5311$). The eigenvalues $E$ calculated from Eqs. (\ref{E-3}) and (\ref{E-4}) are the same as those obtained from the exact diagonalization of the Markov transition matrix $\mathcal{M}$.}\label{Lambda3Lambda4-M1}
\end{table}
\begin{table}[htbp]
  \centering
\begin{tabular}{|c|c|c|c|c|c|} 
\hline\hline $\mu_1$ & $\nu_1$ & $\nu_2$  & $E$  \\ \hline 
 $0.6984-0.1104\ir$ &--&--&  $-2.5526$  \\ 
  $0.6497-0.2790\ir$ &--&--&  $-1.7676$  \\ 
  $0.0614+0.7044\ir$ &--&--&  $-1.0575$  \\ 
  $0.2897-0.0000\ir$ &--&--&  $-0.6133$  \\  
  $0.2284-0.0000\ir$ &--&--&  $-0.7549$  \\ 
-- & $0.6773+0.2031\ir$  &   $0.7020-0.0847\ir$ &  $-2.8832$   \\ 
 -- & $0.5332+0.4644\ir$  &   $0.7024+0.0812\ir$ &  $-2.1364$   \\ 
 -- & {Solution in (\ref{string})} &   {Solution in (\ref{string})} &  ~~~$0.0000$   \\ 
 -- & $0.2828-0.6481\ir$  &   $12.8949-0.0000\ir$ &  $-0.1638$   \\ 
 -- & $0.7030+0.0762\ir$  &   $10.6140+0.0000\ir$ &  $-1.7543$   \\ 
 -- & $0.6817+0.1877\ir$  &   $0.5411-0.4552\ir$ &  $-1.6196$   \\ 
 -- & $0.3089+0.3514\ir$  &   $0.7056+0.8026\ir$ &  $-0.3684$   \\ 
 -- & $0.0466-0.0000\ir$  &   $0.6862+0.1708\ir$ &  $-1.2874$   \\ 
 -- & $0.8613$  &   $1.1611$ &  $-1.0468$   \\ 
 -- & $0.6304+0.3204\ir$  &   $11.0615-0.0000\ir$ &  $-0.6793$   \\ 
 -- & $0.4698+0.1718\ir$  &   $0.9388+0.3433\ir$ &  $-0.8067$   \\ 
 \hline\hline\end{tabular} 
  \caption{Solutions of BAEs (\ref{BAE-3}) and (\ref{BAE-4}) for $N=4, \widetilde{M}=1, M=2, q=0.5, \alpha=0.23, \beta=0.32, \gamma=0.17$ ($a=2.7974,b=2.5891,c=-0.4836,d=-0.5710$).}\label{Lambda3Lambda4-M2}
\end{table}
\begin{table}[htbp]
 \centering
\begin{tabular}{|c|c|c|c|c|c|} 
\hline\hline $\nu_1$ & $\nu_2$ & $\nu_3$ &  $E$   \\ \hline 
$1.6804-0.4682\ir$  &  $0.6695-0.1645\ir$  &  $0.7090+0.0707\ir$ & $-2.9014+0.0703\ir$   \\ 
 $0.7043-0.1730\ir$  &  $0.6982+0.0697\ir$  &  $0.2761-0.0769\ir$ & $-2.9014-0.0703\ir$   \\ 
{Solution in (\ref{string})}  &  {Solution in (\ref{string})}  &  {Solution in (\ref{string})} & ~~~$0.0000$   \\ 
 $0.0532+0.0490\ir$  &  $0.3896-0.5901\ir$  &  $0.0532-0.0490\ir$ & $-0.1963$   \\ 
 $0.1870-0.0000\ir$  &  $0.7037+0.0690\ir$  &  $0.5976-0.3780\ir$ & $-2.2253$   \\ 
 $0.3930-0.2946\ir$  &  $0.1415-0.0000\ir$  &  $0.3930+0.2946\ir$ & $-0.4493$   \\ 
 $4.3284-4.1062\ir$  &  $0.6413-0.2978\ir$  &  $0.0608-0.0577\ir$ & $-0.7280$   \\ 
 $0.5249+0.1373\ir$  &  $0.4457-0.1166\ir$  &  $0.2093+0.1119\ir$ & $-0.8037+0.0679\ir$   \\ 
 $0.8915+0.2332\ir$  &  $1.0499-0.2747\ir$  &  $1.8581+0.9935\ir$ & $-0.8037-0.0679\ir$   \\ 
 $0.1680+0.1227\ir$  &  $0.3877+0.0016\ir$  &  $0.7754+0.0033\ir$ & $-1.0997+0.1109\ir$   \\ 
 $0.1680-0.1227\ir$  &  $0.3877-0.0016\ir$  &  $0.7754-0.0033\ir$ & $-1.0997-0.1109\ir$   \\ 
 $4.1702+4.0169\ir$  &  $4.1702-4.0169\ir$  &  $0.6896+0.1562\ir$ & $-1.3411$   \\ 
 $0.6718-0.0000\ir$  &  $0.6435-0.2931\ir$  &  $1.3394-0.0000\ir$ & $-1.9846$   \\ 
  $4.1231-3.9934\ir$  &  $0.7040-0.0663\ir$  &  $4.1231+3.9934\ir$ & $-1.7716$   \\
 $0.5993+0.3753\ir$  &  $2.8148-0.0000\ir$  &  $0.6870+0.1673\ir$ & $-1.7648$   \\ 
 --& -- & -- & $-1.1529$ \\
 \hline\hline\end{tabular} 
  \caption{Solutions of BAEs (\ref{BAE-4}) for $N=4, M=3, q=0.5, \alpha=0.23, \beta=0.32, \gamma=0.17$ ($a=2.7974,b=1.1511,c=-0.4836,
d=-0.6421$). The $T-Q$ relation in (\ref{TQ-3}) gives the remaining eigenvalue $E=-1.1529$.}\label{Lambda3Lambda4-M3}
\end{table}
\vfill
\begin{table}[htbp]
  \centering
\begin{tabular}{|c|c|c|c|c|c|} 
\hline\hline $\nu_1$ & $\nu_2$ & $\nu_3$ & $\nu_4$ & $E$   \\ \hline 
$0.3211+0.1298\ir$  &  $-0.0017-0.8776\ir$  &  $0.7271+0.0084\ir$  &  $0.6570-0.1524\ir$ & $-3.2814+0.1934\ir$   \\ 
 $0.3211-0.1298\ir$  &  $-0.0017+0.8776\ir$  &  $0.7271-0.0084\ir$  &  $0.6570+0.1524\ir$ & $-3.2814-0.1934\ir$   \\ 
 {Solution in (\ref{string})}  &  {Solution in (\ref{string})}  &  {Solution in (\ref{string})} & {Solution in (\ref{string})}& ~~~$0.0000$   \\ 
 $1.5779-2.8221\ir$  &  $4.3501+0.0000\ir$  &  $1.5779+2.8221\ir$  &  $0.4753+0.5236\ir$ & $-0.2400$   \\ 
 $0.0574-0.7553\ir$  &  $0.6592+0.3317\ir$  &  $1.5888-0.2133\ir$  &  $0.6799-0.0045\ir$ & $-2.4314+0.0716\ir$   \\ 
 $0.0574+0.7553\ir$  &  $0.6592-0.3317\ir$  &  $1.5888+0.2133\ir$  &  $0.6799+0.0045\ir$ & $-2.4314-0.0716\ir$   \\ 
 $0.8513+0.5881\ir$  &  $0.3213+0.6299\ir$  &  $2.3196+0.0000\ir$  &  $0.8513-0.5881\ir$ & $-0.5524$   \\ 
 $3.7321-0.0000\ir$  &  $0.0793+0.1603\ir$  &  $0.6534+0.2703\ir$  &  $0.0793-0.1603\ir$ & $-0.8034$   \\
 $0.0901-1.0542\ir$  &  $1.5430-0.7324\ir$  &  $1.0936+0.2566\ir$  &  $0.8667-0.2035\ir$ & $-0.9090+0.1710\ir$   \\ 
 $1.0936-0.2566\ir$  &  $0.0901+1.0542\ir$  &  $0.8667+0.2035\ir$  &  $1.5430+0.7324\ir$ & $-0.9090-0.1710\ir$   \\ 
 $0.6237-0.3331\ir$  &  $0.2777+0.0000\ir$  &  $0.6947-0.1320\ir$  &  $0.0957+0.7006\ir$ & $-2.0050$   \\  
 $0.1410-0.0000\ir$  &  $0.0771+0.1660\ir$  &  $1.1509+2.4781\ir$  &  $0.6814-0.0000\ir$ & $-1.9055$   \\ 
 $1.8012-0.9756\ir$  &  $1.4831-0.0085\ir$  &  $0.0436-1.3395\ir$  &  $0.6729-0.0003\ir$ & $-1.4569+0.2057\ir$   \\ 
 $1.8012+0.9756\ir$  &  $1.4831+0.0085\ir$  &  $0.0436+1.3395\ir$  &  $0.6729+0.0003\ir$ & $-1.4569-0.2057\ir$   \\ 
 $1.5049-0.0000\ir$  &  $1.0602-1.1811\ir$  &  $0.2104-0.2344\ir$  &  $0.7420+0.0000\ir$ & $-1.5670$   \\ 
 $1.1700-2.4853\ir$  &  $0.6950+0.1302\ir$  &  $1.1700+2.4853\ir$  &  $3.5943+0.0000\ir$ & $-1.4564$   \\ 
 \hline\hline\end{tabular} 
  \caption{Solutions of BAEs  (\ref{BAE-4}) for $N=M=4, q=0.5, \alpha=0.23, \beta=0.32, \gamma=0.17$ ($a=2.7974,b=0.4974,c=-0.4836,d=-0.7429$).}\label{Lambda3Lambda4-M4}
\end{table}

\end{document}